\documentclass[aps,prl,twocolumn,preprintnumbers,amsmath,amssymb,superscriptaddress]{revtex4-1}
\usepackage{graphicx}
\usepackage{subfigure}
\usepackage{epsfig}
\usepackage{dcolumn}
\usepackage{bm}
\usepackage{ulem}
\usepackage{color}

\def\avg#1{\left\langle#1\right\rangle}

\def\be{\begin{equation}}       \def\ee{\end{equation}}
\def\bea{\begin{eqnarray}}      \def\eea{\end{eqnarray}}
\def\ba{\begin{array}}
\def\ea{\end{array}}
\def\bnum{\begin{enumerate} }
\def\enum{\end{enumerate}}

\def\nn{\nonumber}
\def\pa{\partial}
\def\=>{\Rightarrow}
\def\>{\rightarrow}
\def\A{\uparrow}
\def\V{\downarrow}

\def\eye2{Fathbb{I}}

\def\Eq#1{Eq.~(\ref{#1})}
\def\Fig#1{Fig.~\ref{#1}}

\renewcommand{\>}{\rangle}

\begin{document}

\title{Emergent space-time supersymmetry in 3D Weyl semimetals and 2D Dirac semimetals}

\author{Shao-Kai Jian}
\affiliation{Institute for Advanced Study, Tsinghua University, Beijing, 100084, China}
\author{Yi-Fan Jiang}
\affiliation{Institute for Advanced Study, Tsinghua University, Beijing, 100084, China}
\affiliation{Department of Physics, Stanford University, Stanford, CA 94305, USA}
\author{Hong Yao}
\email{yaohong@tsinghua.edu.cn}
\affiliation{Institute for Advanced Study, Tsinghua University, Beijing, 100084, China}

\begin{abstract}
Supersymmetry (SUSY) 
interchanges bosons and fermions but no direct evidences of it have been revealed in nature yet.  In this letter, we observe that fluctuating pair density waves (PDW) consist of {\it two} complex order parameters which can be superpartners of the unavoidably-doubled Weyl fermions in three-dimensional lattice models. We construct explicit fermionic lattice models featuring 3D Weyl fermions and show that PDW is the leading instability via a continuous phase transition as short-range interactions exceed a critical value. Using renormalization group, we theoretically show that ${\cal N}=2$ space-time SUSY emerges at the continuous PDW transitions in 3D Weyl semimetals, which we believe is the {\it first} realization of emergent (3+1)D space-time SUSY in microscopic lattice models. We further discuss possible routes to realize such lattice models and experimental signatures of emergent SUSY at the PDW criticality.
\end{abstract}
\date{\today}

\maketitle
		
\textbf{Introduction:} About four decades ago, the concept of space-time supersymmetry (SUSY) was proposed as an elegant and promising way to solve many fundamental issues in nature such as the hierarchy problem\cite{Weinbergbook,Gervais-Sakita1971, Wess-Zumino1974, Dimopoulos-Georgi1981} and the cosmological constant problem\cite{Cremmer1983}. Despite considerable efforts including recent experiments at the LHC, evidences of SUSY and/or its spontaneous breaking in particle physics are still not definitive. In the mean time, looking for SUSY as an emergent symmetry in condensed matter systems has attracted increasing attentions by mainly asking ``under what circumstances SUSY emerges at low energy and long distances even though it is not respected by constituents at microscopic level?''

In this Letter, we theoretically show that SUSY can emerge in Weyl semimetals in (3+1)D and Dirac semimetals in (2+1)D when they undergo a quantum phase transition into Fulde-Ferrell-Larkin-Ovchinnikov (FFLO) or pair density wave (PDW) phases although their microscopic models are not supersymmetric. Why do we consider transitions into PDW \cite{FF,LO,Berg2007,Berg2009,Radzihovsky2009,Zhai2010,Moore2012,Vakek2014,PALee2014,Fradkin2014,Joseph2014} instead of uniform superconductivity (SC)? Normally, Weyl semimetals in (3+1)D \cite{Nielsen-Ninomiya1983,XGWan2011,Burkov2011,GangXu2011} or Dirac semimetals in (2+1)D have two or more two-component fermions. To be possibly supersymmetric at the transition, equal number of complex bosons (fluctuations of order parameters) and fermions are needed. PDW is a superconducting phase which could have two or more complex order parameters at finite momenta, making it a promising arena to look for emergent SUSY.

We first construct lattice models which feature Weyl semimetals in (3+1)D or Dirac semimetals in (2+1)D and then employ self-consistent mean-field calculations to obtain the phase diagram of these microscopic models (see below) as a function of short-range interactions. Quantum critical points between PDW and semimetals are obtained. Furthermore, we perform renormalization group analysis to demonstrate the emergence of ${\cal N}=2$ SUSY at the PDW criticality of (3+1)D Weyl semimetals with {\it two} Weyl points. For (2+1)D Dirac semimetals, we show that  the ${\cal N}=2$ SUSY emerges at the PDW criticality only there are {\it two} massless Dirac fermions.

It was known that SUSY may emerge at critical points\cite{Shenker1984, Nayak1998, Fendley2003, SungSik2007, YYu-KYang2008, YYu-KYang2010, Fendley2008, Berg2013, Ashvin2014, SungSik2014, Berg2014,Herbut2009}. Previous works are mainly focused on emergent SUSY in low-dimensional (effectively (1+1)D or (2+1)D) systems. Prototype examples of them include the (1+1)D tricritical Ising model where SUSY emerges at the tricritical point\cite{Shenker1984} and the (2+1)D surface states of topological insulators\cite{Hasan-Kane2010,XLQi-SCZhang2011} at transitions into uniform SC\cite{Ashvin2014,SungSik2014,Roy2013}. We emphasize that our work here shows that emergent space-time SUSY could also occur in certain (3+1)D condensed matter systems. We not only construct explicit microscopic lattice models of electrons which are Weyl semimetals and support the desired PDW criticality but also show that the space-time SUSY indeed emerges at the PDW criticality in such (3+1)D Weyl semimetals by tuning a single parameter (strength of onsite interactions). We further stress that the low energy theory at the PDW criticality in (3+1)D Weyl semimetals and (2+1)D Dirac semimetals can support full space-time SUSY instead of the limited SUSY in time direction only, which may lead to important experimental consequences as discussed below.

\textbf{3D Weyl semimetals and PDW instability:} We consider the following microscopic model of spin-1/2 fermions on the cubic lattice at half filling:
\bea
H &=& H_0 + H_1, \label{model4D} \\
H_0 &=& \sum_{\vec k} c^\dag_{\vec k} \Big[\lambda ( \sin k_x \sigma^y - \sin k_y \sigma^x) + d_z(\vec k) \sigma^z \Big] c_{\vec k}
\label{non_int},~~\\
H_1&=&-U\sum_i c^\dag_{i\A}c_{i\A}c^\dag_{i\V}c_{i\V},
\eea	
where $c^\dag_{i\sigma}$ creates an electron on-site $i$ with spin polarization $\sigma=\A,\V$, $\sigma^\alpha$ is a Pauli matrix with spin indices, $d_z(\vec k)=M-2t_1(\cos k_x+ \cos k_y)-4t_2\cos k_x\cos k_y-2t_3[\cos(2k_x)+\cos(2k_y)]-2t_z \cos k_z$, and $H_1$ describes the on-site Hubbard interactions. Even though both time-reversal symmetry (TRS) ${\cal T}$ and inversion symmetry are explicitly broken, the Hamiltonian respects ${\cal T}_x'={\cal T}M_x$ and ${\cal T}_y'={\cal T}M_y$, where $M_\alpha$ represent mirror operation reflecting the $\alpha$-axis. It is clear that with appropriate parameters in $H_0$ the two non-degenerate bands touch at only {\it two} discrete momenta $\pm \vec K\equiv(0,0,\pm k^\ast_z)$, $k^\ast_z=\cos^{-1} [\frac{M-4t_1-4t_2-4t_3}{2t_z}]$, which are generally incommensurate and which are so called ``Weyl points''. The degeneracy of two Weyl points is protected by the symmetry ${\cal T}'_x$ or ${\cal T}'_y$ such that at half-filling only two Weyl points cross the Fermi level. By expanding the Bloch Hamiltonian around $\pm \vec K$ up to linear order in $\vec p= \vec k\pm \vec K$, we obtain the low-energy Hamiltonian of two Weyl fermions:
\bea
H_{0,\textrm{eff}} \!=\! \sum_{\vec p} \psi^\dag_{\vec p} [ (v_{fx} p_x \sigma^x + v_{fy} p_y \sigma^y) +v_{fz} p_z \sigma^z\tau^z] \psi_{\vec p},~~~
\eea
where $v_{fx} = v_{fy} =\lambda$ and $v_{fz} = 2t_z\sin k_z^\ast$. Here, $\tau^\alpha$ is a Pauli matrix with valley indices ($\pm$).

\begin{figure}						
\centering
\includegraphics[width=7.5cm]{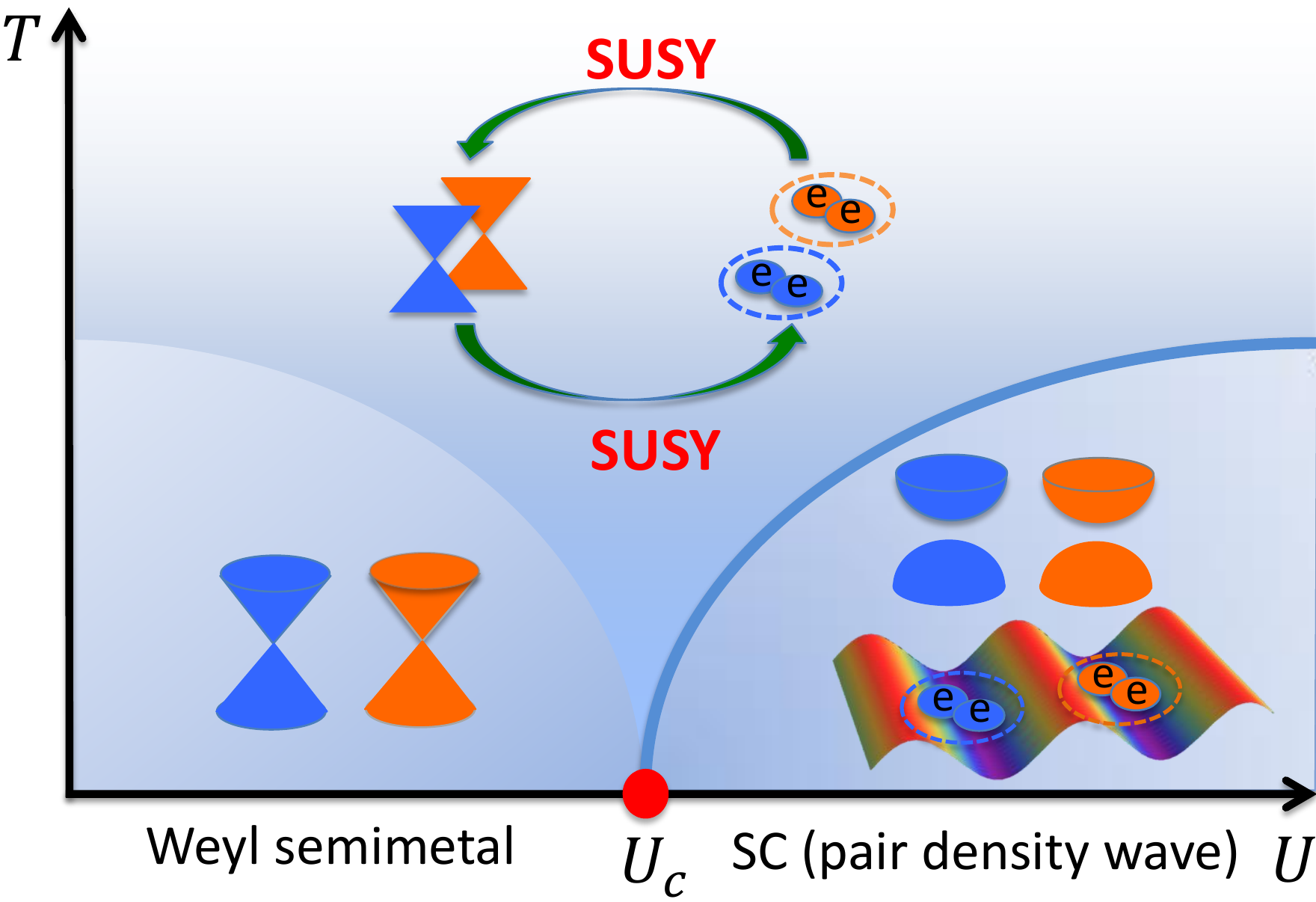}
\caption{The schematic phase diagram of the 3D lattice model with two Weyl fermions as a function of the Hubbard attraction $U$. 
At the PDW critical point $U=U_c$, the ${\cal N}=2$ space-time SUSY in (3+1)D emerges in low energy and long distance.}
\label{pd-weyl}
\end{figure}

Now, we are ready to consider how interactions affect the low energy physics of the Weyl fermions. Due to the vanishing density of states at Weyl points, weak short-range interactions do not have qualitative effect on Weyl fermions as they are irrelevant in RG. However, for sufficiently strong attractions Weyl fermions are generically unstable to broken symmetry phases. For onsite attractions, superconductivity and charge density waves (CDW) are two leading choices for broken symmetry phases. Assuming that phase transitions originate from instability of low-energy Weyl fermions, there are only three possible broken symmetry phases induced by onsite attractions: uniform SC ($\avg{\psi_+\sigma^y\psi_-}\neq 0$), PDW ($\avg{\psi_+\sigma^y\psi_+}\neq 0$ and $\avg{\psi_-\sigma^y\psi_-}\neq 0$), and $2\vec K$-CDW ($\langle\psi^\dag_+\psi_-\rangle\neq0$). It is straightforward to see that both $2\vec K$-CDW and uniform SC cannot fully gap out the Weyl fermions while even infinitesimal PDW ordering can fully gap out the two Weyl fermions. Consequently, we expect that Weyl fermions are unstable to PDW phases as the interaction exceeds a threshold when low-energy Weyl fermions are responsible for the putative instability \cite{footnote-strong}.

We investigate the phase diagram as a function of $U$ by performing self-consistent mean-field (MF) calculations. By setting $t_1=1.0$, $t_2=-1.5$, $t_3=0.41$, $t_z=1.0$, $\lambda=1.0$, and $M=0.16$, the Weyl points are at $\pm \vec K=(0,0,\pm k^\ast_z)$ with $k^\ast_z=\frac{5\pi}{12}$. We obtain the phase diagram as shown in \Fig{pd-weyl}. The continuous phase transition from the Weyl semimetal to the PDW phase occurs at $U=U_c\approx 8.1$. For $U_c<U<U'_c\approx 10.7$, the PDW phase is characterized by $\avg{\psi_+\sigma^y\psi_+} = \avg{\psi_-\sigma^y\psi_-}\neq 0$. At $U=U'_c$, the system undergoes a first-order phase transition into the commensurate $(\pi,\pi,\pi)$-CDW phase which could be understood in the limit of strong coupling: with strong $U$, electrons tend to form onsite pairs which organize them into a $(\pi,\pi,\pi)$-CDW pattern by reducing nearest-neighbor repulsion energy between pairs.

\textbf{Emergent (3+1)D SUSY at PDW criticality:}  We have shown above that PDW is the leading instability of the Weyl semimetal as the interaction is varied across a critical value and the phase transition is continuous. Close to the PDW quantum critical point, the low energy physics can be captured by two Weyl fermions $\psi_\pm$ at $\pm\vec K$ and two pairing order parameters $\phi_\pm$ at $\pm \vec{Q}=\pm 2\vec{K}$. The effective field theory near the PDW criticality reads
\bea
S &=& S_{f}+ S_{b} + S_I, \label{effaction4D} \\
S_{f} &=& \int d^4x \sum_{n=\pm} \Big[ \psi_n^\dag \partial_\tau \psi_n+\sum_{j=1}^3 i v_{fj}\psi_n^\dag \gamma_n^j \partial_j \psi_n\Big],  \\
S_{b} &=& \int d^4x \bigg\{ \sum_{n=\pm} \Big[|\partial_\tau \phi_n|^2 + \sum_{j=1}^3 v_{bj}^2 |\partial_j \phi_n|^2   \nn\\
&& ~~~~~~~~~ +r|\phi_n|^2 +u |\phi_n|^4 \Big]+u_{+-} |\phi_+|^2 |\phi_-|^2 \bigg\},  \\
S_I &=&  \int d^4x g \sum_{n=\pm} \Big[ \phi_n \psi_n \sigma^y \psi_n + h.c.\Big],
\eea	
where $\gamma_{\pm}^j=(\sigma^x,\sigma^y,\pm \sigma^z)$ depending on whether the Weyl fermion at valley $n$ is right/left-handed. Here $S_f$ is the same as non-interacting Weyl fermions since four-fermion interaction term is irrelevant in the sense of RG and $S_b$ describes the fluctuation of PDW order parameters $\phi_n$ close to the quantum critical point. Note that the term $\phi^\ast_n \pa_\tau\phi_n$ does not appear in $S_b$ because of particle-hole symmetry\cite{footnote-PHS}. Fermion velocities are in general anisotropic, {\it i.e.} $v_{fx} \ne v_{fy} \ne v_{fz} $ unless certain symmetries impose full or partial isotropy. Consequently, boson velocities are anisotropic as well, {\it i.e.} $v_{bx}\ne v_{by}\ne v_{bz}$. To be supersymmetric at low energy, among other requirements fermions and bosons should have identical velocities, namely emergent Lorentz symmetry. In $S_b$ we consider four-boson terms at most since higher order terms are irrelevant. $S_I$ describes the Yukawa coupling between fermions and bosons. As the energy density should be bounded from below, we require $u>0$ and $u_{+-}\ge -2u$. Thus, the effective action $S$ above is of most general form near the PDW phase transition in Weyl semimetals with two Weyl fermions.

\begin{figure}[t]						
\includegraphics[width=4.cm]{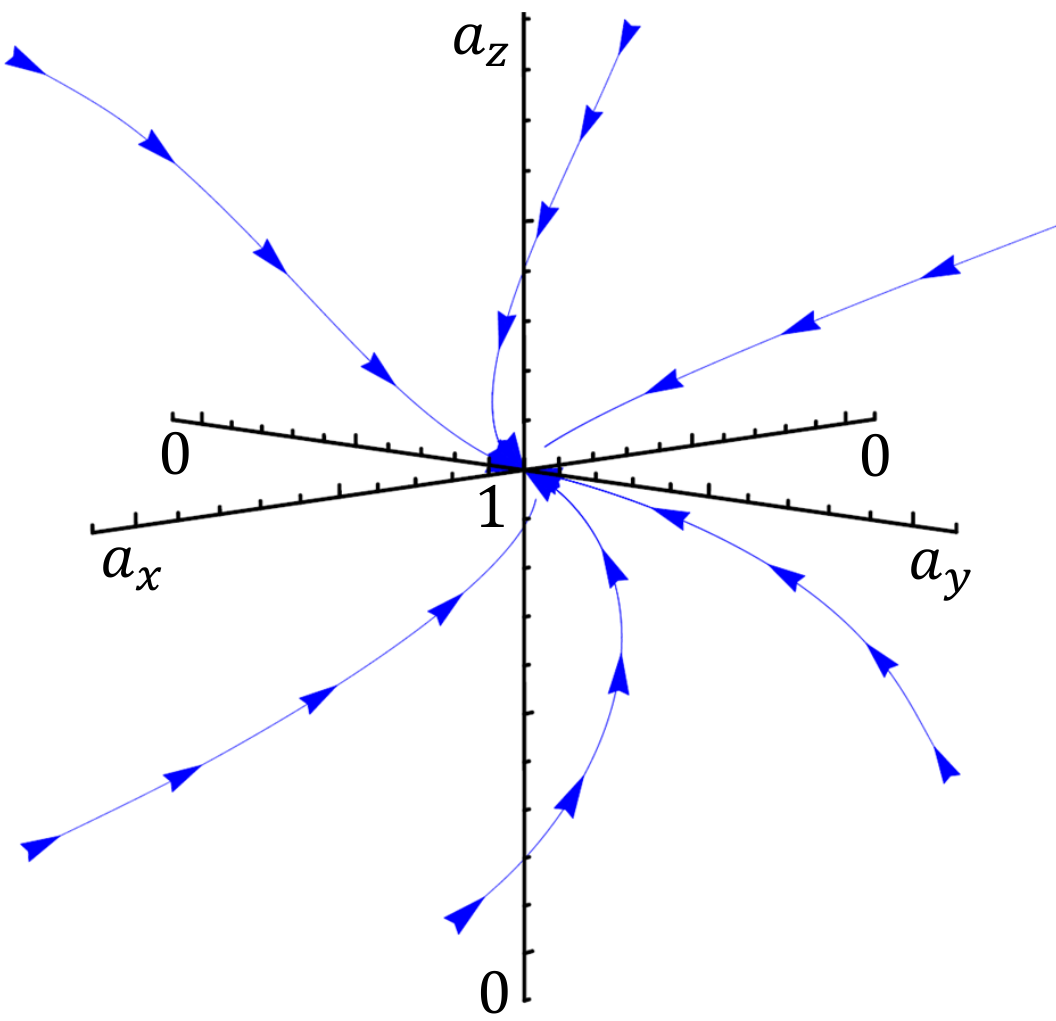}~~~~~~~
\includegraphics[width=3.7cm]{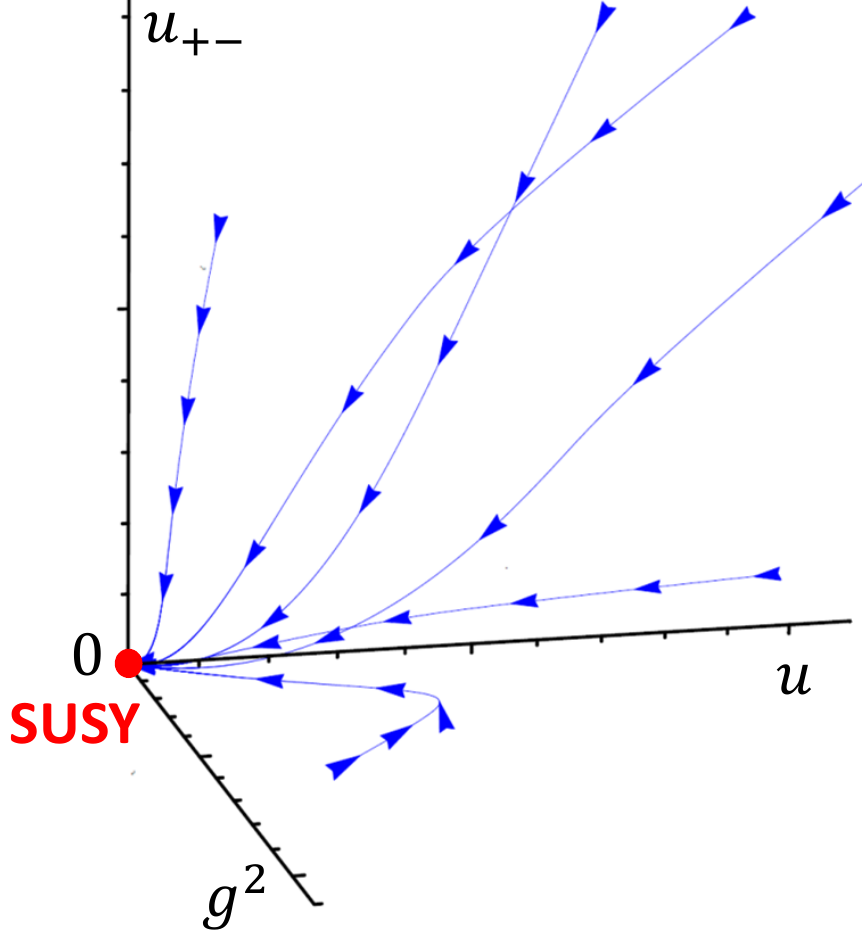}
\caption{(a) The RG flow of velocity ratios $a_j$ between bosons and fermions for (3+1)D Weyl fermions at the PDW critical point. (b) The RG flow of interaction parameters $g^2$, $u$ and $u_{+-}$. It is clear that $g$, $u$ and $u_{+-}$ are all marginally irrelevant.  }
\label{rgflow4D}
\end{figure}

We perform one-loop RG analysis in $D=4-\epsilon$ dimensions for the effective action \Eq{effaction4D} at criticality ($r=0$). Physical (3+1)D systems correspond to $\epsilon=0$. By performing Wilsonian RG of gradually integrating out fast modes between the cutoff $\Lambda$ and $\Lambda(l)=\Lambda e^{-l}$, we obtain the following RG flow equations of boson/fermion velocities and their ratios:
\bea
&&\frac{d a_j}{dl} =\frac{g^2}{8\pi^2 v_{fz}^3} \frac{1-a_j^2}{w_1w_2 a_j} -\frac{g^2}{4\pi^3 v_{fz}^3} a_j (F_j-F_0), \\
&&\frac{d w_{j}}{dl}=\frac{g^2}{4\pi^3 v_{fz}^3} w_j (F_j-F_3),
\eea
where $a_j\equiv v_{bj}/v_{fj}$ ($j$=$x,y,z$) is the ratio of the boson to fermion velocity and $w_{j}=v_{fj}/v_{fz}$ ($j$=$x,y$) characterizes velocity anisotropy of fermions. Here $F_j$ are functions of $a_j$ and $w_j$, as given in Supplemental Material \cite{supplemental}. When $a_j=1$, $F_1=F_2=F_3=F_0=\frac{\pi}{w_xw_y}$. It is clear that there is a plane of fixed points, $a^\ast_{x}=a^\ast_y=a^\ast_z=1$ with arbitrary $w^\ast_{x}$ and $w^\ast_{y}$, in the $(a_j,w_{x},w_{y})$-hyperspace. The flow of $a_j$ towards to the fixed point $a^\ast_j=1$  which is showed in \Fig{rgflow4D}(a). By linearizing the RG equations in the vicinity of an arbitrary point in the fixed plane, we find that the scaling field away from the fixed plane is negative while within the fixed plane it is zero, which shows that this fixed plane is stable (see the Supplemental Material\cite{supplemenal}).
		
Once $a_j,w_{x},w_{y}$ flow into the fixed plane, fermions and bosons have identical velocities in low energy: $v_{fj}=v_{bj}$. By spatial rescaling, we can set $v_{bj} = v_{fj} = 1$, and perform the RG analysis of coupling constants. The RG flow equations are given by  				
\bea
\frac{d g^2}{dl} &=&\epsilon g^2 -\frac{3}{2} g^4 , \\
\frac{d u}{dl} &=& \epsilon u+ 2g^4 - \frac{5}{2} u^2 -\frac{1}{8} u_{+-}^2 - g^2 u , \\ 		
\frac{d u_{+-}}{dl} &=& \epsilon u_{+-} -2 u u_{+-} - \frac{1}{2} u_{+-}^2 -g^2 u_{+-},
\eea	
where $g^2\to \frac{g^2}{2\pi^2}$, $u\to \frac{u}{2\pi^2}$, and $u_{+-}\to \frac{u_{+-}}{2\pi^2}$ were assumed implicitly. For Weyl fermions in (3+1)D, {\it i.e.} $\epsilon=0$,  there is only one {\it stable} fixed point, $(g^*,u^*,u_{+-}^*)=(0,0,0)$, which corresponds to free fermions and bosons. In the low energy limit, the interaction terms turn out to be marginally irrelevant and their values flow to zero logarithmically: $g^2\sim\frac{2}{3\log(\Lambda/E)}$, $u\sim \frac{2}{3\log(\Lambda/E)}$, and $u_{+-}\sim \frac{1}{\log^2(\Lambda/E)}$, where $\Lambda$ represents an energy cutoff of order of the band width and $E$ represents a probing energy scale. It is interesting to notice that $u$ and $g^2$ flow to zero at a fixed ratio of 1. Moreover, the velocity ratio between bosons and fermions flows to 1 even slower: $a_j-1\sim \frac{1}{\log\log(\Lambda/E)}$. At the fixed point, we obtain the effective action: 		
\bea
S_\textrm{3+1D-SUSY}=\sum_{n=\pm} \int d^4x (\partial^\mu \phi_n^* \partial_\mu \phi_n -i \psi_n^\dag \bar{\sigma}_n^\mu \partial_\mu \psi_n).~~~~~
\eea
It is clear that this action describes two independent copies of free boson and fermions which respect the ${\cal N}=2$ SUSY in (3+1)D. Indeed, it is invariant under supersymmetry variation: $\delta \phi_n = \sqrt{2} \epsilon_n \psi_n$ and $
\delta \psi_n = i \sqrt{2} \sigma^\mu \bar{\epsilon}_n \partial_\mu \phi_n$,
where $\epsilon_n$ is a left-handed Weyl spinor and its complex conjugate $\bar{\epsilon}_n$ is a right-handed Weyl spinor both parameterizing the infinitesimal variations of supersymmetry transformation. Therefore, we have shown that full space-time SUSY emerges at the PDW phase transition of (3+1)D Weyl semimetals. This is one of central results of the present paper.

We have shown emergence of SUSY at the PDW criticality of Weyl semimetals with two Weyl fermions, which break TRS explicitly. For time-reversal invariant but non-centrosymmetric Weyl semimetals, there are at least four Weyl points. For a generic Weyl semimetal with $N_f$ ($N_f\ge 4$) Weyl fermions, the charge-4e Josephson coupling is present. If it is marginally irrelevant, the ${\cal N}=2$ SUSY in (3+1)D emerges similarly. It is also interesting to note that it is possible to realize a Weyl semimetal with $N_f=1$ as on the surface of a 4+1D Chern insulator\cite{SCZhang2001}; there the ${\cal N}=2$ SUSY emerges at a putative transition into a uniform superconducting state.

\begin{figure}[t]						
\includegraphics[width=4.cm]{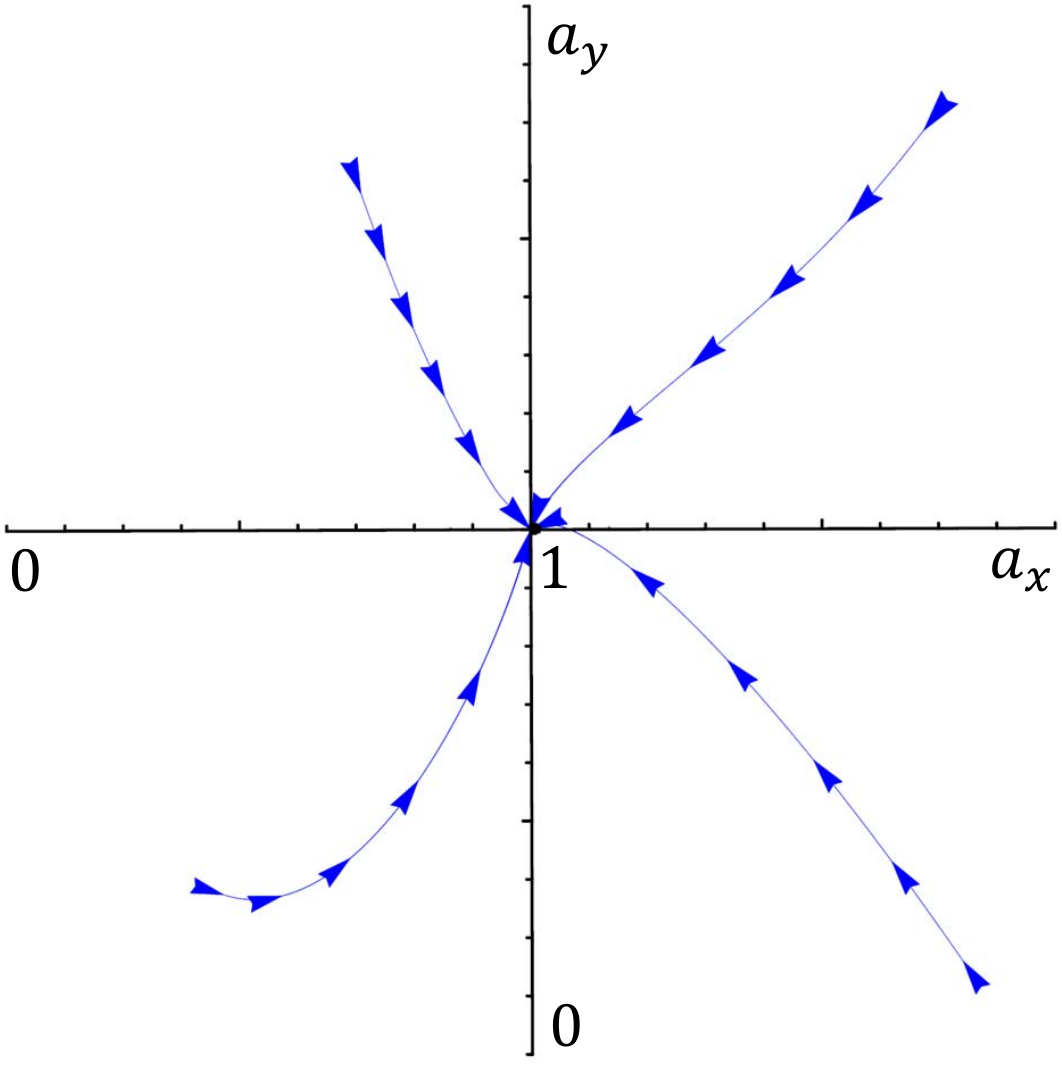}~~~~
\includegraphics[width=4.2cm]{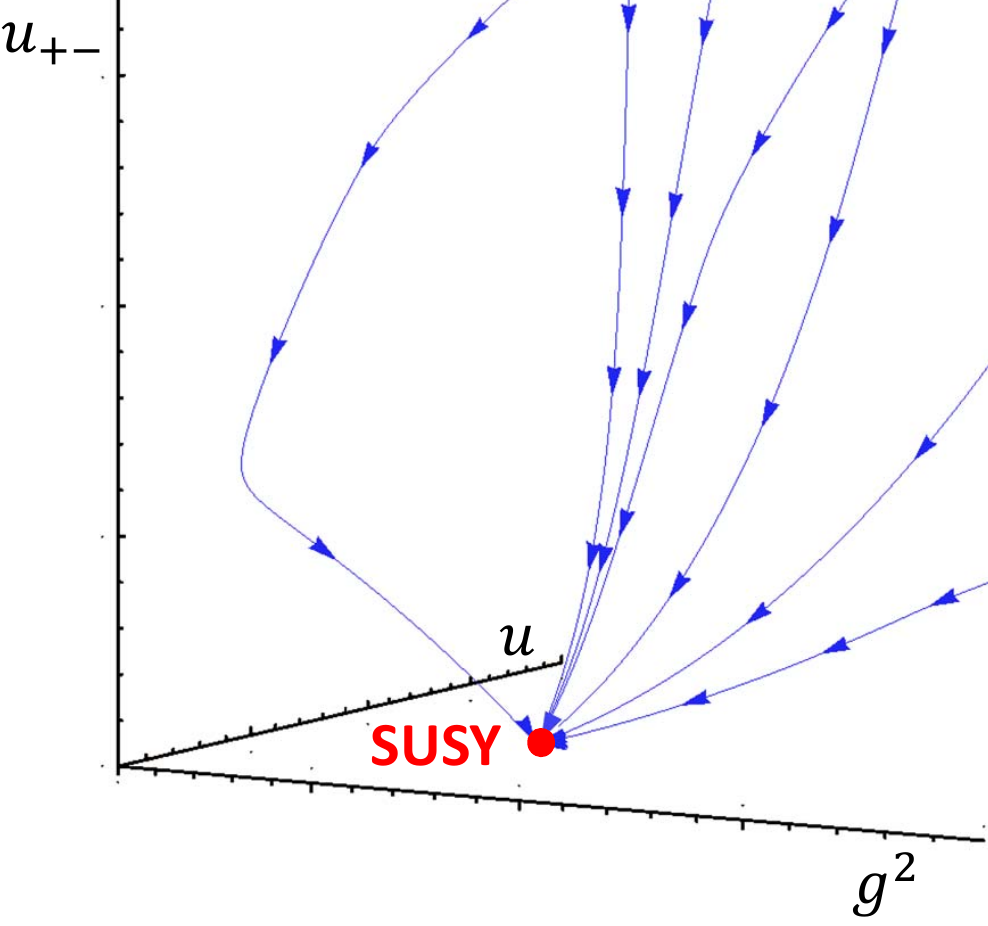}
\caption{(a) The RG flow of velocity ratios $a_j$ between bosons and fermions in (2+1)D Dirac semimetals at the PDW critical point. (b) The RG flow of interaction parameters $g$, $u$ and $u_{+-}$. It is clear that $u_{+-}$ are all marginally irrelevant while $g$ and $u$ flow to the supersymmetric point. }
\label{rgflow3D}
\end{figure}

\textbf{SUSY in 2D Dirac semimetals:} At PDW transitions in (2+1)D Dirac semimetals, the counting of bosonic and fermionic modes is similar to the one in (3+1)D Weyl semimetals, we expect that emergent SUSY should also occur at PDW transitions in (2+1)D Dirac semimetals, where pairing is between two fermions from the same Dirac node. The effective action at the PDW transition in (2+1)D semimetals with {\it two} massless Dirac fermions is the same as \Eq{effaction4D} except it is one dimension lower. Since the RG flow equations are derived using dimensional expansion in $\epsilon=4-D$, we can obtain the RG flow equations in the (2+1)D by setting $\epsilon=1$. Indeed, the velocities of bosons and fermions flow to the same value in low energy, as shown in \Fig{rgflow3D}. At the fixed point, $u^\ast_{+-}=0$, indicating that the PDW order parameters at opposite momenta decouple in low energy, and $u^\ast=(g^\ast)^2 \neq 0$. Consequently we conclude that two copies of ${\cal N}=2$ SUSY may emerge at the putative PDW transition of 2D Dirac semimetals\cite{SungSik2007}. How about the PDW transition in Dirac semimetals with four or more Dirac points? In Supplemental Material\cite{supplemental}, we show that SUSY does {\it not} emerge in this case because the charge-4e Josephson couplings which relate the phases of various pairing order parameters are marginally relevant and tend to lock the relative phases between them in low energy.

\begin{figure}[t]						
\centering
\includegraphics[width=5.8cm]{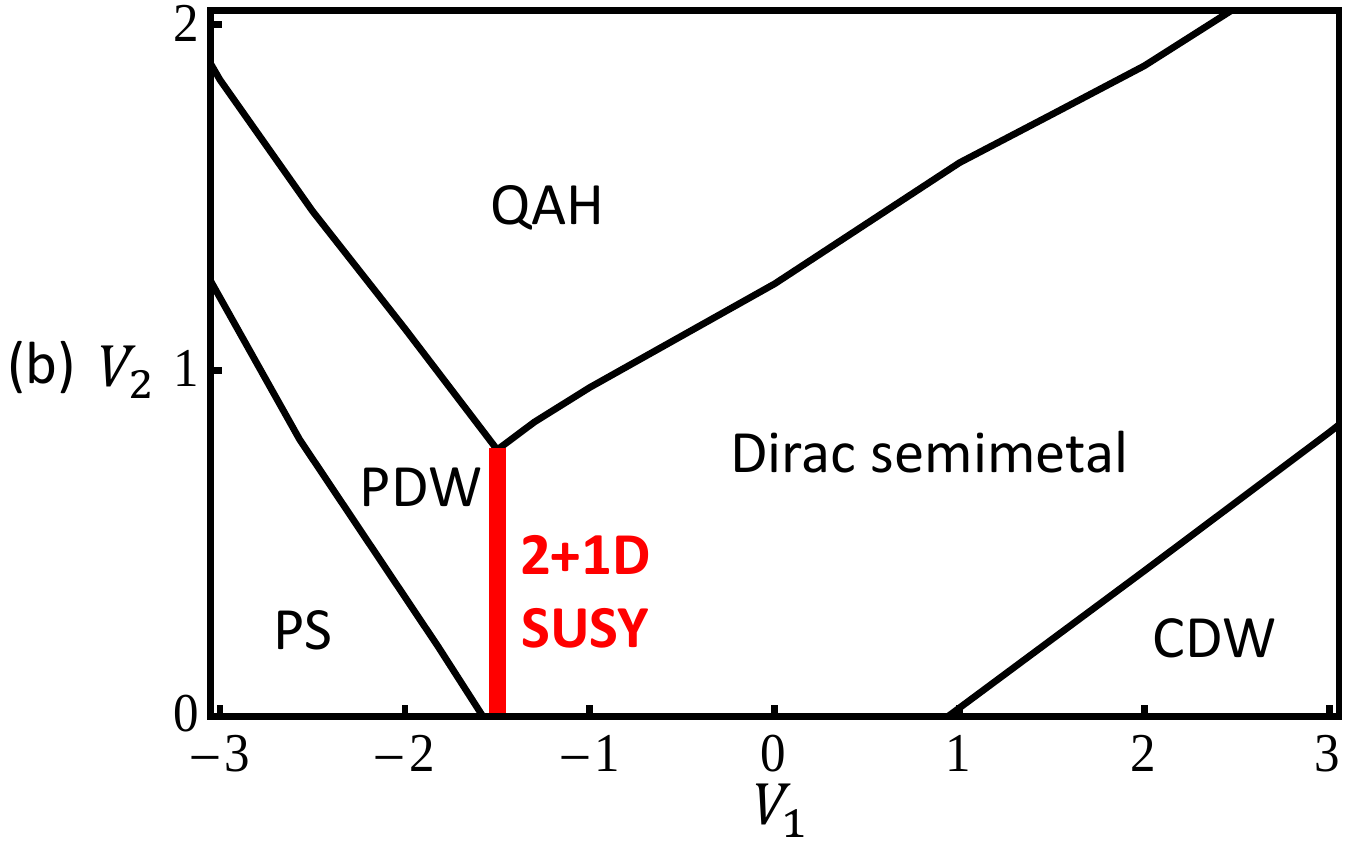}
\caption{The mean-field phase diagram of the spinless fermion $t$-$V_1$-$V_2$ honeycomb model as a function of $V_1$ and $V_2$. Here QAH and CDW label quantum anomalous Hall and sublattice-CDW, respectively; PS represents phase separation. For this model (and also another model studied in the Supplemental Material \cite{supplemental}), at the PDW criticality the ${\cal N}=2$ space-time SUSY emerges in low energy.}
\label{pd-dirac}
\end{figure}

Here, we consider a {\it spinless} fermion model on the honeycomb lattice with short-range density interactions and show that a continuous phase transition between the Dirac semimetals and PDW phases occurs by varying interaction parameters. The Hamiltonian is given by
\bea\label{model3D2}
H=\sum_{\avg{ij}} (-tc^\dag_ic_j+H.c.)+V_1\sum_{\avg{ij}}  n_i n_j +V_2\sum_{\avg{\avg{ij}}}  n_i n_j, ~~~~~
\eea
where $V_1$ and $V_2$ are NN and NNN interactions. The band dispersion with NN hopping features two Dirac points at $\pm \vec K$. The quantum phase diagram at half-filling is obtained by mean-field calculations, as shown in \Fig{pd-dirac}. It was known that this model supports a quantum anomalous Hall phase within certain range of repulsive $V_2$\cite{Raghu2009}. We show that the PDW phase where pairing occurs between two fermions from the same valley $\vec K$ or $-\vec K$ is realized in certain region of attractive $V_1$ and repulsive $V_2$; the quantum phase transition is second-order. Accordingly to the RG analysis above, two copies of ${\cal N}=2$ SUSY emerges at the phase boundary between the Dirac semimetals and the PDW phase in this model.

\textbf{Consequences of emergent SUSY:} SUSY is an intrinsically fermionic symmetry which rotates fermions into their bosonic superpartners. Consequently, universal critical exponents of fermions and corresponding bosons at criticality are related nontrivially. For instance, the scaling dimensions of $\psi$ and $\phi$ satisfy $\Delta_\psi=\Delta_\phi+1/2$. At the PDW critical point of (3+1)D Weyl semimetals, SUSY of free fermions and bosons with identical velocities emerges for which the critical exponents are known exactly: $\Delta_\psi=\Delta_\phi+1/2=3/2$. Even though the emergent SUSY is between free fermions and bosons with identical velocity, it has nontrivial consequence in the nearby broken symmetry phase. For the (2+1)D Dirac semimetals at the PDW transition, even though the RG flow are analyzed in the one-loop level with introducing dimensional expansion parameter $\epsilon=4-D$, the emergent (2+1)D ${\cal N}=2$ SUSY enables one to determine exactly the scaling dimensions $\Delta_\psi=\Delta_\phi+1/2=7/6$ and anomoulous dimensions $\eta_\psi=\eta_\phi=1/3$ \cite{Aharony1997}. Consequently, fermion local density of states at the SUSY critical point is $\rho(\omega)\propto |\omega|^{\frac{4}{3}}$ which may be measured by STM experiments. Moreover, slightly into the PDW phase, we obtain $m_\psi=m_\phi\propto (r-r_c)^\nu$ with $\nu=3/5$. The fermion mass or superconducting gap here can be measured by tunneling experiments such as STM. For both (3+1)D and (2+1)D ${\cal N}=2$ SUSY quantum critical points discussed above, the exact field scaling dimensions and other critical exponents may be verified in future numerical simulations of the microscopic models in (3+1)D [\Eq{model4D}] and in (2+1)D [\Eq{model3D2}].

\textbf{Concluding remarks:} We have shown that emergent ${\cal N}=2$ space-time SUSY occurs in (3+1)D Weyl semimetals with {\it two} Weyl points and in (2+1)D Dirac semimetals with {\it two} Dirac points when these systems are tuned to PDW phase transitions by varying short-range interactions. Emergent SUSY has nontrivial consequences such as letting us obtain exact scaling dimensions of fermion and boson fields as well as other critical exponents. One remaining but important issue is how to possibly realize them in nature. One promising way is to employ ultracold atoms loaded into an optical lattice, where relatively strong onsite attractions can be achieved by tuning the system close to Fechbach resonance. For the honeycomb $t$-$V_1$-$V_2$ model, we may consider dipolar molecules with electric dipoles or atoms with magnetic dipoles \cite{review} loaded into an optical honeycomb lattice. Because of the angular dependence of dipolar interactions, attractive $V_1$ and repulsive $V_2$ may be achieved by polarizing the dipoles on different sublattices into opposite directions which are out-of-plane.

{\it Acknowledgement}: We are indebted to Eduardo Fradkin, Sung-Sik Lee, T. Senthil, Kun Yang, and Hui Zhai for helpful discussions. This work is supported in part by the National Thousand-Young-Talents Program (HY) and by the NSFC under Grant No. 11474175 at Tsinghua University (SKJ, YFJ, and HY).

\begin{widetext}
\section{Supplementary Material}
\renewcommand{\theequation}{S\arabic{equation}}
\setcounter{equation}{0}
\renewcommand{\thefigure}{S\arabic{figure}}
\setcounter{figure}{0}

\subsection{A. The quantum phase diagrams by mean-field analysis}
In the main text, we have introduced two microscopic lattice models which feature either a Weyl semimetal phase in three spatial dimensions or a Dirac semimetal phase in two spatial dimensions in the non-interacting limit. Here, we construct another 2D lattice model that has {\it two} massless Dirac fermions and that also feature a PDW quantum phase transition:
\bea
H = \sum_{\vec k} c^\dag_{\vec k} \Big[\lambda \sin k_x \sigma^y  + d_z \sigma^z \Big] c_{\vec k}-U\sum_i c^\dag_{i\A}c_{i\A}c^\dag_{i\V}c_{i\V},~~~~~
\eea
where $d_z(\vec k)=M-2t_1\cos k_x-2t_2\cos(2k_x)-2 t_y\cos k_y $. This model respects ${\cal T}'_x={\cal T}M_x$ which protects the massless Dirac dispersion at $(0,\pm k^\ast_y)$ with $k^\ast_y=\cos^{-1}[\frac{M-2t_1-2t_2}{2t_y}]$. The energy degeneracy of the two Dirac points is protected by the symmetry ${\cal T}'_x$ such that at half-filling only these Dirac points cross the Fermi level. For $\lambda=1.0, t_1=1.0, t_2=-1.2, t_y=1.0, M=0.2$, $k^\ast_y=\frac{5\pi}{12}$; we obtain the quantum phase diagram as shown in main text. The PDW is at momenta $\pm \vec Q=\pm (0,\frac{5\pi}{6})$ and the second-order phase transition occurs at $U=U_c\approx 6.0$. At the PDW criticality, two copies of ${\cal N}=2$ SUSY emerge in low energy. Because of the vanishing density of states at the Weyl/Dirac points, weak short-range interactions do not qualitatively change the semimetal phase. In other words, short-range interactions are irrelevant in the RG sense as long as they are weak. However, broken symmetry phases should occur when interactions are strong enough.

We employ mean-field analysis to obtain the quantum phase diagram as a function of interactions and to determine the nature of the quantum phase transitions between the semimetal phase and broken symmetry phases. For the interactions considered, our mean-field analysis found that the PDW ordering with pairing of fermions within the same valley is the leading instability out of the semimetal phase in all three microscopic models. As shown in \Fig{pdwop}, the quantum phase transitions between the semimetal and the PDW phases are all continuous, which justifies the low energy effective field theory description of such critical points.

\begin{figure}[h]						
\centering
\subfigure[]{\includegraphics[width=5.9cm]{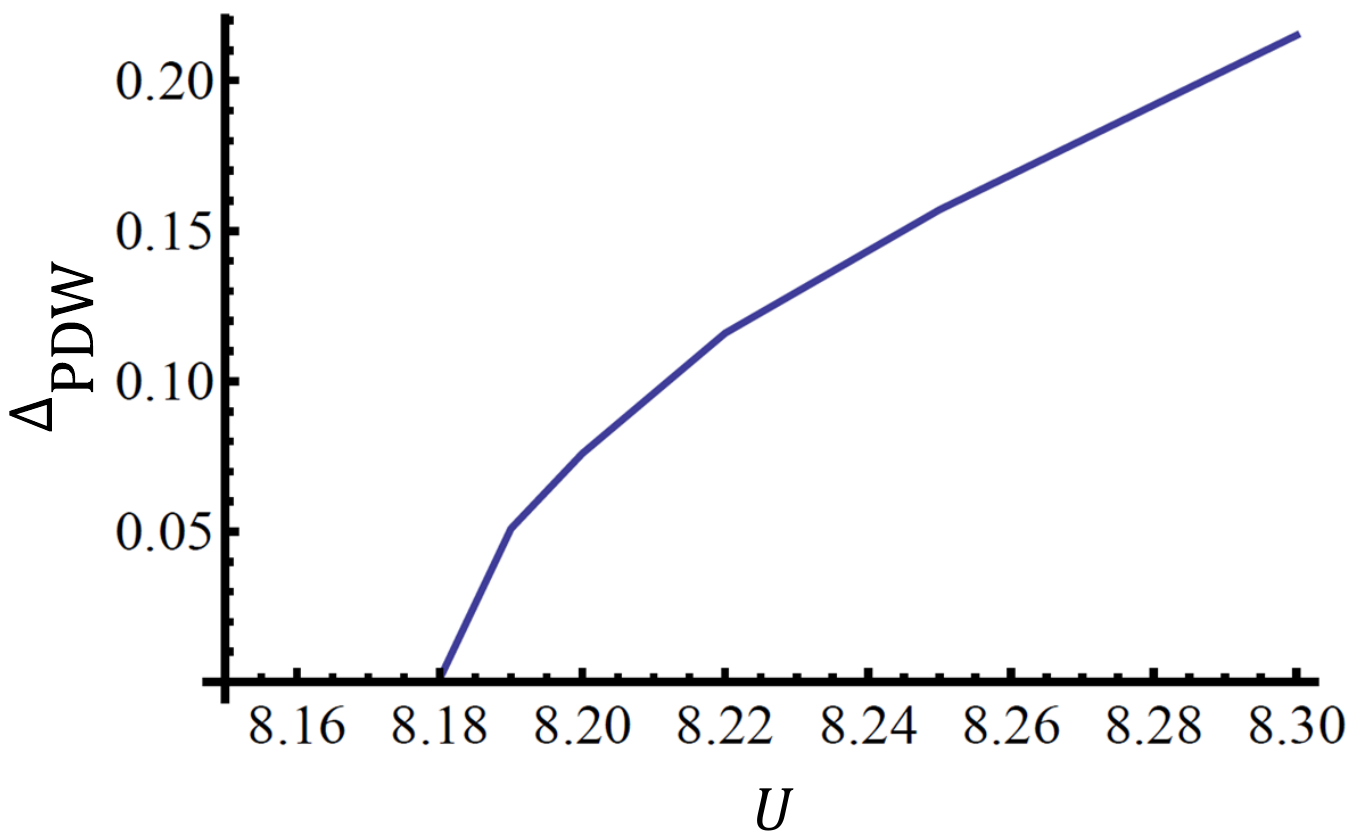}}
\subfigure[]{\includegraphics[width=5.9cm]{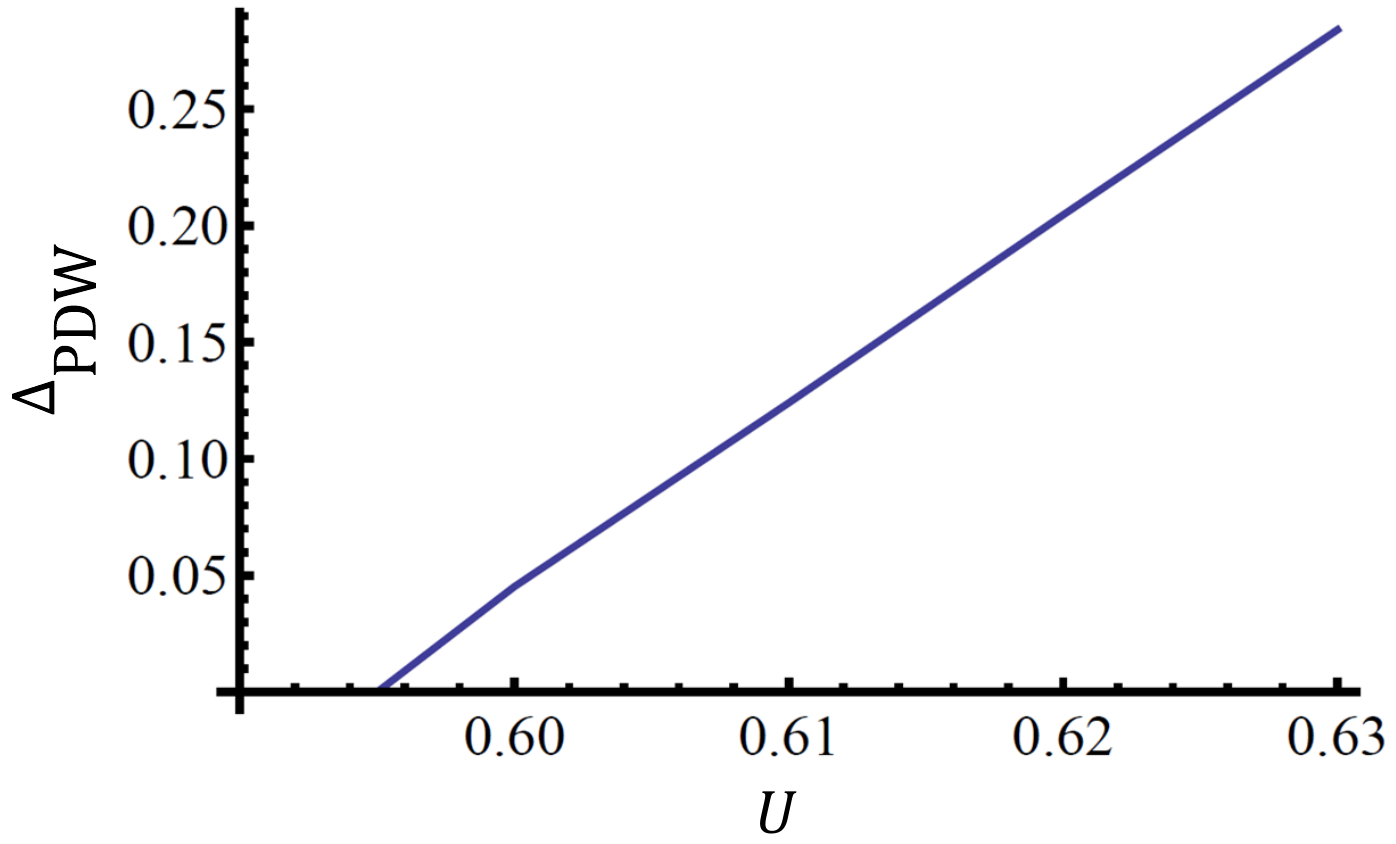}}
\subfigure[]{\includegraphics[width=5.9cm]{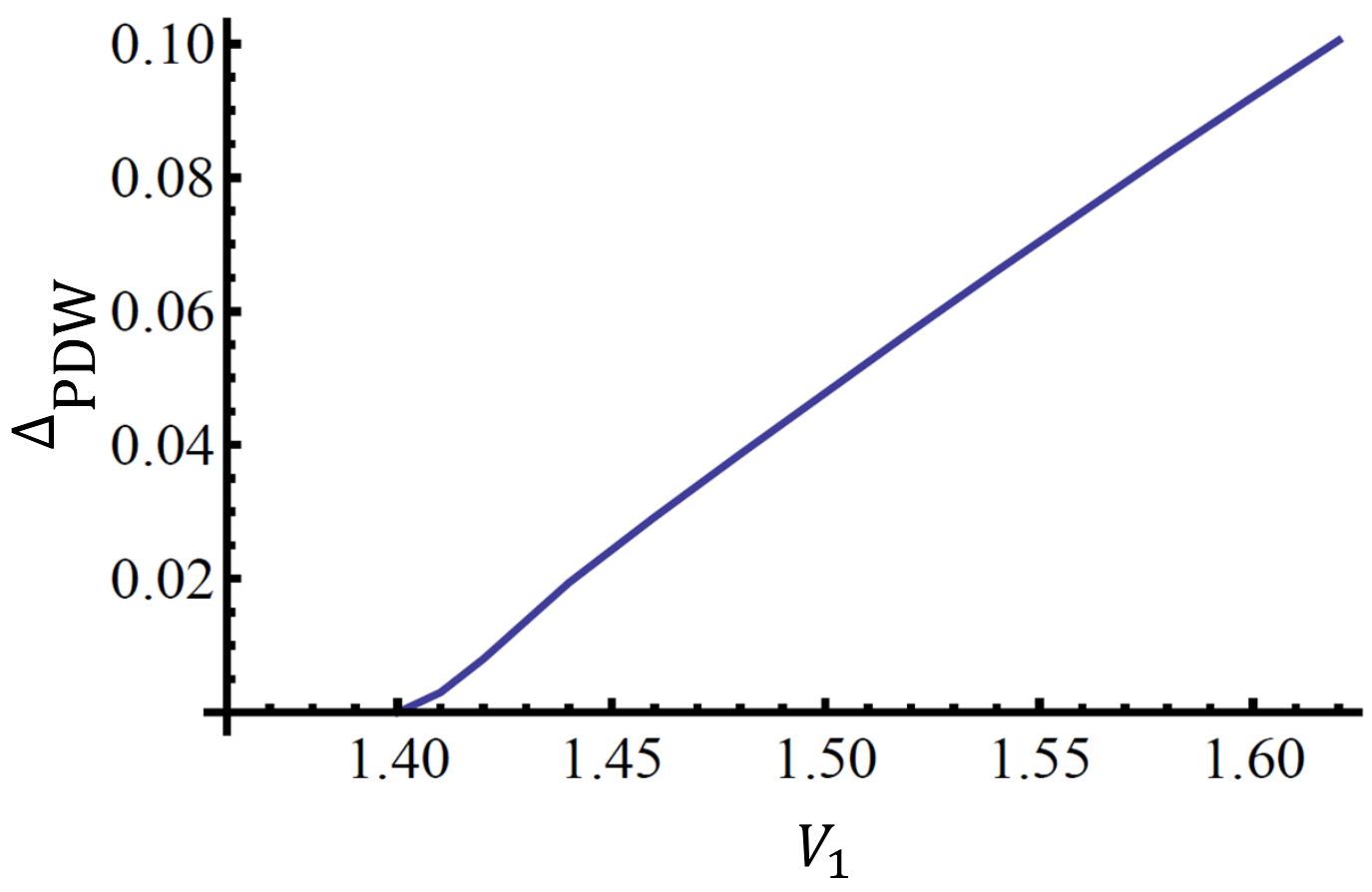}}
\caption{The quantum phase transitions between the PDW and semimetal phases are shown to be all continuous in the three microscopic models. (a) The PDW order parameter $\Delta_\textrm{PDW}$ as a function of $U$ in the cubic lattice model [Eq.(1) in the main text] with two Weyl fermions. (b) The PDW order parameter $\Delta_\textrm{PDW}$ as a function of $U$ in the square lattice model [Eq.(S1)] with two Dirac fermions. (c) The PDW order parameter $\Delta_\textrm{PDW}$ as a function of $V_1$ (fixing $V_2=0.5$) in the honeycomb lattice model [Eq.(15) in the main text] with two Dirac fermions.}
\label{pdwop}
\end{figure}

\subsection{B. The RG analysis for velocity flows at PDW transitions in 2+1D and 3+1D: emergent Lorentz symmetry}

At the PDW criticality $r=0$ of Weyl semimetals in 3+1D or Dirac semimetals in 2+1D, the RG flow of velocity ratios between bosons and fermoins does not depend on the number of nodes (at least in one-loop level). We obtain the following RG flow equations of velocity for generic case identical to the ones with two nodes:
\begin{eqnarray}
\frac{d a_j}{dl} =\frac{g^2}{8\pi^2 v_{fz}^3} \frac{1-a_j^2}{w_1w_2 a_j} -\frac{g^2}{4\pi^3 v_{fz}^3} a_j (F_j- F_0),~~~~~~~~
\frac{d w_{j}}{dl}=\frac{g^2}{4\pi^3 v_{fz}^3} w_j (F_j-F_3),
\end{eqnarray}
where the functions $F_j$ are given by
\bea
F_1 &=& a_1^2 w_1^2 \int_{-1}^1 d\cos\theta \int_0^{2\pi}d\phi \frac{\sin^2 \theta \cos^2 \phi (\sqrt{A}+2 \sqrt{B})}{\sqrt{AB^3}(\sqrt{A}+\sqrt{B})^2}, \\
F_2 &=& a_2^2 w_2^2  \int_{-1}^1 d\cos\theta \int_0^{2\pi}d\phi \frac{\sin^2 \theta \sin^2 \phi (\sqrt{A}+2 \sqrt{B})}{\sqrt{AB^3}(\sqrt{A}+\sqrt{B})^2}, \\
F_3 &=& a_3^2 \int_{-1}^1 d\cos\theta \int_0^{2\pi}d\phi \frac{ \cos^2 \theta (\sqrt{A}+2 \sqrt{B})}{\sqrt{AB^3}(\sqrt{A}+\sqrt{B})^2},
\eea
\bea
F_0 &=& \int_{-1}^1 d\cos\theta \int_0^{2\pi}d\phi \frac{1}{(\sqrt{A}+ \sqrt{B})^2 \sqrt{B}},
\eea
with $A= w_1^2 \sin^2 \theta \cos^2 \phi+w_2^2 \sin^2 \theta \sin^2 \phi + \cos^2 \theta$ and
$B=  a_1^2 w_1^2 \sin^2 \theta \cos^2 \phi+a_2^2 w_2^2 \sin^2 \theta \sin^2 \phi + a_3^2\cos^2 \theta$. It is clear that the velocity ratios between bosons and fermions have fixed point values of 1; namely bosons and fermions have identical velocities in low energy. However, the velocity anisotropy $w_j>0$ are marginal. In this fixed plane, these integral $F_j$ can be evaluated analytically $F_j= \frac{\pi}{w_1 w_2}$ for $j=0,x,y,z$. We further show that this fixed plane is stable. The linearized RG equations at any point in this plane are given by
\begin{eqnarray}
\frac{d}{dl}\left( \begin{array}{ccccc} \delta a_1 \\ \delta a_2 \\ \delta a_3 \\ \delta w_1 \\ \delta w_2 \end{array}\right) = \frac{5g^2}{6 w^*_1 w^*_2} \left( \begin{array}{ccccc} -1 & 0 & 0 & 0 & 0 \\ 0 & -1 & 0 & 0 & 0  \\ 0 & 0 & -1 & 0 & 0 \\ \frac{2w^*_1}{5} & 0 & -\frac{2w^*_1}{5} & 0 & 0\\ 0 &\frac{2w^*_2}{5}  & -\frac{2w^*_2}{5} & 0 & 0 \end{array}\right) \left( \begin{array}{ccccc} \delta a_1 \\ \delta a_2 \\ \delta a_3 \\ \delta w_1 \\ \delta w_2 \end{array}\right).
\end{eqnarray}
Diagonalizing the scaling matrix above, we find that all $a_j-1$ are irrelevant while $w_j$ are marginal. This demonstrates that the fixed plane is stable.

\subsection{C. The RG analysis for $N_f$ ($N_f\ge 4$) 2+1D massless Dirac fermions at PDW transitions: no emergent SUSY}
For 2+1D Dirac semimetals with $N_f$ massless Dirac fermions, we assume that all the Dirac points have degenerate energy required by symmetries of the system in question. At the PDW transition, the low-energy effective action close to the PDW criticality is given by
\begin{eqnarray}
S &=& S_{f}+ S_{b} + S_I, \label{effaction4DNf} \\
S_{f} &=& \int d^3x \sum_{n=1}^{N_f} \Big[ \psi_n^\dag \partial_\tau \psi_n+\sum_{j=1}^3 i v_{fj}\psi_n^\dag \gamma_n^j \partial_j \psi_n\Big],  \\
S_{b} &=& \int d^3x \bigg\{ \sum_{n=1}^{N_f} \Big[|\partial_\tau \phi_n|^2 \!+\! \sum_{j=1}^3 v_{bj}^2 |\partial_j \phi_n|^2 +r|\phi_n|^2 +u |\phi_n|^4 \Big] \nn\\
&&~~~~~~~~~~~+ \sum_{n \ne n'}^{N_f} \frac{u_{nn'}}{2} |\phi_n|^2 |\phi_{n'}|^2 + \sum_{n\ne n'}^{\frac{N_f}{2}} \frac{v_{nn'}}{2} (\phi_n\phi_{\bar{n}} \phi_{n'}^\ast\phi^\ast_{\bar{n'}}+c.c.) \bigg\}, ~ \\
S_I &=&  \int d^3x g \sum_{n=1}^{N_f} \Big[ \phi_n \psi_n \psi_n + \phi_n^* \psi_n^\dag \psi_n^\dag\Big],
\end{eqnarray}
where $\bar n\equiv n\pm \frac{N_f}{2}$ labels the massless Dirac fermion having momentum opposite to $n$ and $v_{nn'}$ term describes the charge-4e Josephson coupling between two {\it different} pairs of condensates $n\bar n$ and $n'\bar {n'}$, which are absent when there are only {\it two} massless Dirac fermions.  Here $u_{nn'}=u_{n'n}$ and $v_{nn'}=v_{n'n}$ due to notation redundancy. Because all the Weyl fermions are connected by symmetries of the system, $u_{nn'}=u_{\bar n\bar{n'}}$ and all $u_{n\bar n}$ are identical. The RG flow equations of velocities of bosons and fermions do not depend on the number of Dirac points; consequently, bosons and fermions have identical velocities in low energy as obtained obtained above and Lorentz symmetry emerges.

Now let's investigate the RG flows of coupling constants. Especially, flows of the new charge-4e Josephson coupling term is crucial to whether SUSY emerges. For simplicity, we consider the simplest case of $N_f=4$ and the four Dirac points are labelled by $1,2,3\equiv\bar 1,4\equiv \bar 2$. The degeneracy of the four Dirac points is ensured by symmetry.  In this simple case, there are six independent coupling parameters: $g$, $u$, $u_{12}$, $u_{13}\equiv u_{1\bar 1}$, $u_{14}$, and $v_{12}$, for which the RG equations in $D=4-\epsilon$ read
\begin{eqnarray}
\frac{dg^2}{dl} &=& \epsilon g^2 - \frac{3}{2} g^4,\\
\frac{du_1}{dl} &=& \epsilon u_1- g^2 u_1+2 g^4 -\frac{1}{8}(20u_1^2+u_{13}^2+u_{12}^2+u_{14}^2),\\
\frac{du_{13}}{dl} &=& \epsilon u_{13}- g^2 u_{13} - \frac{1}{8}(4u_{13}^2 + 4u_{12}u_{14}  + 16u_1u_{13}+ 2v_{12}^2) ,\\
 \frac{du_{12}}{dl} &=& \epsilon u_{12} - g^2 u_{12}- \frac{1}{8}(4u_{12}^2+4u_{13}u_{14}+16u_1u_{12}+ 2v_{12}^2)  ,\\
\frac{du_{14}}{dl} &=& \epsilon u_{14}-g^2u_{14}-\frac{1}{8}(4u_{14}^2+4u_{13}u_{12}+16u_1 u_{14}+ 2v_{12}^2) ,\\
 \frac{dv_{12}}{dl} &=& \epsilon v_{12} -  g^2 v_{12} - \frac{1}{8} (4 u_{12}+4u_{14}+4u_{13})v_{12},
\end{eqnarray}
for which we find that there is only a fixed point that could maintain the energy density bounded frow below, that is $(g^{*2},u^*,u^*_{1\bar 1},u^*_{12},v^*_{12})=(\frac{2\epsilon}{3},\frac{2\epsilon}{3},0,0,0)$, in 2+1D (namely $\epsilon=1$). However, this fixed point is unstable since it is relevant along $v_{12}$ direction. If we linearize the RG equation in the vicinity of this fixed point, we could see that $\frac{d \delta v_{12}}{dl} =\frac{\epsilon}{3} \delta v_{12}$ where $\delta v_{12} =v_{12}-v^\ast_{12}$. This means that in low energy limit, the charge-4e Josephson coupling is relevant.

Consequently, not all the phases of pairing order parameters could fluctuate independently at the phase transition; in other words, the bosonic degrees of freedom is less than the number of two-component fermions. SUSY does not emerge for $N_f\ge 4$ massless Dirac fermions at PDW transitions in 2+1D even though bosons and fermions have identical velocities in low energy. The ${\cal N}=2$ SUSY emerges at the PDW transition only when there are $N_f=2$ massless Dirac fermions.

\end{widetext}


\begin{thebibliography}{99}
\bibitem{footnote-strong} Because the putative PDW instability is not in the limit of weak coupling, it is possible that phases from the stong-coupling limit such as the $(\pi,\pi,\pi)$-CDW could preempt the PDW transition depending on microscopic details of the model. It is possible to have a direct transition from the Weyl semimetal to the $(\pi,\pi,\pi)$-CDW for certain range of parameters. For instance, when $t_1=1.0$, $t_2=-0.5$, $t_3=-0.53$, $t_z=1.0$, $M=0.38$, and $\lambda=1.0$, we observe a direct transition from Weyl semimetal to CDW occuring at $U\approx 4.7$. The phase transition is continuous.
\end{thebibliography}
\end{document}